\documentclass[reprint,amsmath,amssymb,aps,pre,floatfix]{revtex4-2}
\usepackage{graphicx}
\usepackage{dcolumn}
\usepackage{bm}
\usepackage{bookmark}
\usepackage{color}
\graphicspath{{./},{./figs/}}
\begin{document}

\title{Dynamics of microswimmers near a soft penetrable interface}

\author{Chao Feng$^1$}
\author{John J. Molina$^1$}
\author{Matthew S. Turner$^{1,2}$}
\author{Ryoichi Yamamoto$^1$}
\email{ryoichi@cheme.kyoto-u.ac.jp}
\affiliation{
$^1$Department of Chemical Engineering, Kyoto University, Kyoto 615-8510, Japan\\
$^2$Department of Physics, University of Warwick, Coventry CV4 7AL, UK}

\date{\today}

\begin{abstract}
Few simulations exist for microswimmers near deformable interfaces. Here, we present numerical simulations of the hydrodynamic flows associated with a single microswimmer embedded in a binary fluid mixture. The two fluids demix, separated by a penetrable and deformable interface that we assume to be initially prepared in its planar ground-state. We find that the microswimmer can either penetrate the interface, move parallel to it or bounce back off it. We analyze how the trajectory depends on the swimmer type (pusher/puller) and the angle of incidence with respect to the interface. Our simulations are performed in a system with periodic boundary conditions, corresponding to an infinite array of fluid interfaces. A puller reaches a steady state in which it either swims parallel to the interface or selects a perpendicular orientation, repeatedly penetrating through the interface. In contrast, a pusher follows a bouncing trajectory between two interfaces. We discuss several examples in biology in which swimmers penetrate soft interfaces. Our work can be seen as a highly simplified model of such processes.
\end{abstract}

\maketitle

\section{introduction}

Microswimmers, including flagellated bacteria such as \textit{E. coli} \cite{berg2004coli} and motile, single-celled eukaryotes such as \textit{Chlamydomonas} \cite{drescher2010direct}, are common in biology and usually exist in complex fluid environments in nature.
Systematic studies on the dynamics of microswimmers that can help us to understand their complex behavior will thus allow us to further our understanding of basic biology and provide a guide for developing artificial micromachines. The latter could have great potential in various technological and biomedical applications \cite{elgeti2015physics,lauga2016bacterial}, such as targeted drug delivery \cite{laage2006molecular} or therapies using microrobots \cite{ceylan20183d}.

Most theoretical/simulation studies on the dynamics of microswimmers have focused on their dynamics in homogeneous environments \cite{downton2009simulation,volpe2014simulation,oyama2016purely}.
Among the few works that have focused on the dynamics of microswimmers in inhomogeneous multiple-fluid systems, studies have usually focused on swimmers in the vicinity of solid/fluid interfaces \cite{volpe2011microswimmers,li2014hydrodynamic,ishimoto2017dynamics,fadda2020dynamics} or liquid/gas interfaces \cite{ishimoto2013squirmer}.
These previous studies have revealed that microswimmers can be strongly influenced by liquid/solid and liquid/gas boundaries; they may either loiter near, escape from, or glide along the boundary \cite{li2014hydrodynamic}.
Over a longer time scale, circular motion at the boundary has even been observed, with the swimming orientation determined by the boundary conditions \cite{ishimoto2017dynamics}.

Although biological microswimmers are usually found in complex or inhomogeneous fluids, most studies on their dynamics have focused on swimmers in simple homogeneous host fluids. Studies on inhomogeneous systems with soft and/or penetrable interfaces are still limited in number due to the high associated computational costs \cite{shaik2017motion,daddi2019frequency}.
In the present study, we take a step toward understanding the behavior of microswimmers in complex environments by performing direct numerical simulation (DNS) of swimmers in a binary (Newtonian) fluid mixture. In particular, we fully account for the deformable and penetrable nature of the interface between the two phase-separated fluids.

This paper is organized as follows. First, we present the details of our theoretical model and numerical method. We then provide a comprehensive analysis of our DNS results, in which we observe two distinct motions at the interface, depending on the type of swimmer and the incidence angle: (1) transmission across the interface and (2) bouncing back from the interface. Finally, we also present a detailed analysis and characterization of the resulting steady-state behavior.

\section{simulation methods}

\subsection{The squirmer model}

To model microswimmers, the squirmer model is employed in this work.
It is a widely used model for a self-propelled particle in which the swimmer is simplified as a spherical object with a modified stick boundary condition at its surface \cite{Lighthill,downton2009simulation}.
The slip velocity at the surface of the sphere is
\begin{equation}
    {\bm{u}} ^s( {\bm{\hat{r}}} )=\sum_{n=1}^{\infty} \frac{2}{n(n+1)} B_{n}P^{'}_{n}( \cos \theta) \sin \theta {\bm {\hat{\theta}}}, \label{eq:sq}
\end{equation}
where ${\bm {\hat{r}}} $ is the unit vector directed from the center of the squirmer toward a point on its surface,
${\bm{\hat{\theta}}}$ is the unit polar angle vector at ${\bm{\hat{r}}}$, and
$\theta = \cos^{-1} ({\bm{\hat{r}}} \cdot {\bm{\hat{e}}})$ is the polar angle between ${\bm {\hat{r}}}$ and the swimming direction $\bm {\hat{e}}$.
$P^{'}_{n}$ is the derivative of the Legendre polynomial of the $n$th order,
and $B_{n}$ is the magnitude of the $n$th mode.
Here, radial deformation is ignored;
therefore, this surface velocity has only tangential components and is responsible for the self-propulsion of the swimmer \cite{pak2014generalized,ishikawa2006hydrodynamic}.

In this work, only the first two modes in Eq.~\ref{eq:sq} are retained:
\begin{equation}
    {\bm{u}} ^s( {\theta} )=B_1(\sin \theta + \frac{\beta} {2} \sin {2\theta} ){\bm {\hat{\theta}}}, \label{eq:sq2}
\end{equation}
The coefficient $B_1$ in Eq.~\ref{eq:sq2} is physically related to the steady-state swimming velocity of the squirmer via $\nu_0=2/3B_1$,
and the ratio $\beta=B_2/B_1$ determines the squirmer's swimming type and its strength.
When $\beta$ is negative, the squirmer is a pusher and generates extensile flow fields in the direction of propulsion;
when $\beta$ is positive, the squirmer is a puller generating contractile flow fields.
The marginal case of $\beta=0$ corresponds to a neutral particle that swims with
the potential flow in the surrounding fluid.
Different types of squirmers can be mapped to different kinds of microorganisms in nature.

\subsection{Smoothed profile method} \label{SPM}
To simulate the dynamics of a swimming system while fully considering the hydrodynamic interactions, we employ the smoothed profile (SP) method \cite{yamamoto2021smoothed}.
In this method, all boundaries,
including both fluid/solid and fluid/fluid boundaries,
are considered to possess a finite interfacial thickness $\xi$. This greatly simplifies the modeling and improves the computational efficiency of the method.
Fluid/solid boundaries are implicitly accounted for by introducing a phase field function $\phi({\bm r})$, which is equal to $1$ within solid domains (inside the squirmer particles), is equal to $0$ within the fluid domain (outside of the squirmer particles), and smoothly varies between $0$ and $1$ across the interface. Thus, such an interface can be represented by the gradient of the phase field, which will be nonzero only within the interfacial domains.

A modified (incompressible) Navier--Stokes equation is employed as the governing equation for the total fluid velocity $\bm{u}$:
\begin{align}
        \rho(\partial_t+\bm{u}\cdot\bm{\nabla})\bm{u}  &= -\bm{\nabla}{p}+\nabla\cdot\bm{\sigma}+\rho(\phi\bm{f}_p+\phi\bm{f}_{sq}), 
        \label{eq:NS}\\
        \bm{\nabla}\cdot\bm{u}                         &=0,    
\end{align}
where $\sigma=\eta (\bm{\nabla}\bm{u} + \bm{\nabla}\bm{u}^{\mathrm T})$ is the Newtonian stress tensor (viscosity $\eta$) and $\rho$ is the fluid density. The term $\phi\bm{f}_p$ appearing on the right-hand side of Eq.~\eqref{eq:NS} is introduced to enforce the rigidity of the particles; likewise,
the term $\phi\bm{f}_{sq}$ is introduced to enforce the ``squirming'' boundary condition at the surfaces of the particles (Eq.~\eqref{eq:sq2}).

The total velocity is defined in terms of the fluid velocity field $\bm{u}_f$ and the particle velocity field $\bm{u}_p$ as
\begin{align}
        \bm{u}                                         &=(1-\phi)\bm{u}_f+\phi\bm{u}_p,\\      
        \phi\bm{u}_p                                   &=\sum_i{\phi_i}[\bm{V}_i+\bm{\Omega}_i\times\bm{R}_i],        
\end{align}
where the first term, $(1-\phi)\bm{u}_f$, represents the velocity field of the binary fluid, while the second term, $\phi\bm{u}_p$, represents the particles' velocity field, which is defined in terms of the positions $\bm{R}_i$, velocities $\bm{V}_i$, and angular velocities $\bm{\Omega}_i$ of the particles (where $i$ is the particle index).

The dynamics of the rigid particles are determined by the Newton--Euler equations of motion:
\begin{align}
        \dot{\bm{R}_i} &=\bm{V}_i,      \label{eq:NE}\\
        \dot{\bm{Q}_i} &={\rm skew}(\bm{\Omega}_i)\cdot\bm{Q}_i,     \\
        {M}_p\dot{\bm{V}_i} &=\bm{F}^H_i+\bm{F}^C_i+\bm{F}^{ext}_i, \\
        \bm{I}_p\cdot\dot{\bm{\Omega}_i}  &=\bm{N}^H_i+\bm{N}^{ext}_i,
\end{align}
where $M_p$ and $\bm{I}_p$ are the mass and moment of inertia, respectively, of the particles; $\bm{Q}_i$ is the orientation matrix of particle $i$;
and skew($\bm{\Omega}_i$) is the skew-symmetric matrix of the angular velocity $\bm{\Omega}_i$.
The hydrodynamic forces and torques are given by $\bm{F}^H_i$ and $\bm{N}^H_i$, $\bm{F}^C_i$ represents direct particle--particle interactions ($\bm{N}^C_i = 0$), and
$\bm{F}^{ext}_i$ and $\bm{N}^{ext}_i$ are the external forces and torques, respectively.
Detailed descriptions of the SP method and its implementation can be found in our earlier publications \cite{yamamoto2004smooth,PhysRevE.71.036707,molina2013direct,yamamoto2021smoothed}.

\subsection{Binary fluid model}
The host fluid in our system is modeled as a phase-separating binary fluid mixture using the Cahn--Hilliard (CH) model, which, coupled with the Navier--Stokes hydrodynamics, yields the so-called Model H \cite{arai2020direct,lecrivain2020eulerian,lecrivain2018diffuse,lecrivain2017direct,lecrivain2016direct}. We refer to the two phases of this binary mixture as fluids A and B. The spatial distributions of fluids $A$ and $B$ are given by order parameters $\psi_A ({\bf r})$ and $\psi_B({\bm{r}})$, respectively, with $0 \le \phi_\alpha \le 1$. The composition of the fluid mixture is then determined by the order parameter $\psi(\bm{r})$,
\begin{equation}
        \psi = \psi_A-\psi_B,
        \label{eq:psi}
\end{equation}
which takes a value of $1$ in the $A$ domain and a value of $-1$ the $B$ domain, where the fractions of the constituent components (fluid and particles) must sum to unity:
\begin{equation}
        \psi_A+\psi_B+\phi  =1.
\end{equation}

To account for the binary fluid nature of the host fluid, an additional force term is introduced in Eq.~\ref{eq:NS}:
\begin{eqnarray}
        \rho(\partial_t+\bm{u}\cdot\bm{\nabla})\bm{u} = &-&\bm{\nabla}{p} + \bm{\nabla} \cdot \sigma  - {\psi}\bm{\nabla} \mu_\psi \nonumber \\
        &-& {\phi}\bm{\nabla} \mu_\phi + \rho(\phi\bm{f}_p+\bm{f}_{sq}),
        \label{eq:navier_stokes_model_h}
\end{eqnarray}
where 
$\mu_\psi = \delta F/\delta \psi$
and
$\mu_\phi = \delta F/\delta \phi$
are the locally-defined chemical potentials with respect to $\psi$ and $\phi$, defined as
functional derivatives of the Ginzburg-Landau (GL) free energy $F$.
The time evolution of $\psi$ is given by the following CH equation:
\begin{equation}
    \begin{aligned}
        \frac{\partial \psi}{\partial t} + (\bm{u}\cdot \bm{\nabla}) \psi & = \kappa \nabla ^2 \mu_\psi,
        \label{eq:cahn_hilliard}
    \end{aligned}
\end{equation}
where $\kappa$ is the mobility coefficient. \\
The free energy $F$ can be represented as follows:
\begin{equation}
    \begin{aligned}
        F = f(\psi) + \frac{\alpha}{2} (\bm{\nabla} \psi)^{2} + w \xi \psi (\bm{\nabla}\phi)^{2} .
        \label{eq:freeenergy}
    \end{aligned}
\end{equation}
In Eq.~\ref{eq:freeenergy}, the first term $f( \psi )=\frac{1}{4} \psi^4 - \frac{1}{2} \psi^2$ represents the Landau double-well potential and has two minima at $\psi=1$ and $-1$. 
The second term is the potential energy associated with the fluid $A$/$B$ interface.
The third term represents the particles' affinity for each of the fluid $A$/$B$ phases. 
Thus, the chemical potentials are
\begin{equation}
    \begin{aligned}
        \mu_\psi=f'(\psi)+\alpha\bm{\nabla}^2\psi + w \xi (\bm{\nabla}\phi)^2 ,
        \label{eq:potential_1}
    \end{aligned}
\end{equation}
and
\begin{equation}
    \begin{aligned}
        \mu_\phi= 2w\xi(\bm{\nabla}\psi\bm{\nabla}\phi+\psi\bm{\nabla}^2\phi) .
        \label{eq:potential_2}
    \end{aligned}
\end{equation}
In the present study, to keep the system as simple as possible, we assume that fluids $A$ and $B$ are immiscible but otherwise possess identical physical properties. In addition, we assume that the swimmers interact with the interface only hydrodynamically.
Therefore, we set $w=0$ in the present simulations. 

\section{results}
\begin{figure}[tbh]
    \includegraphics[width=0.7\linewidth]{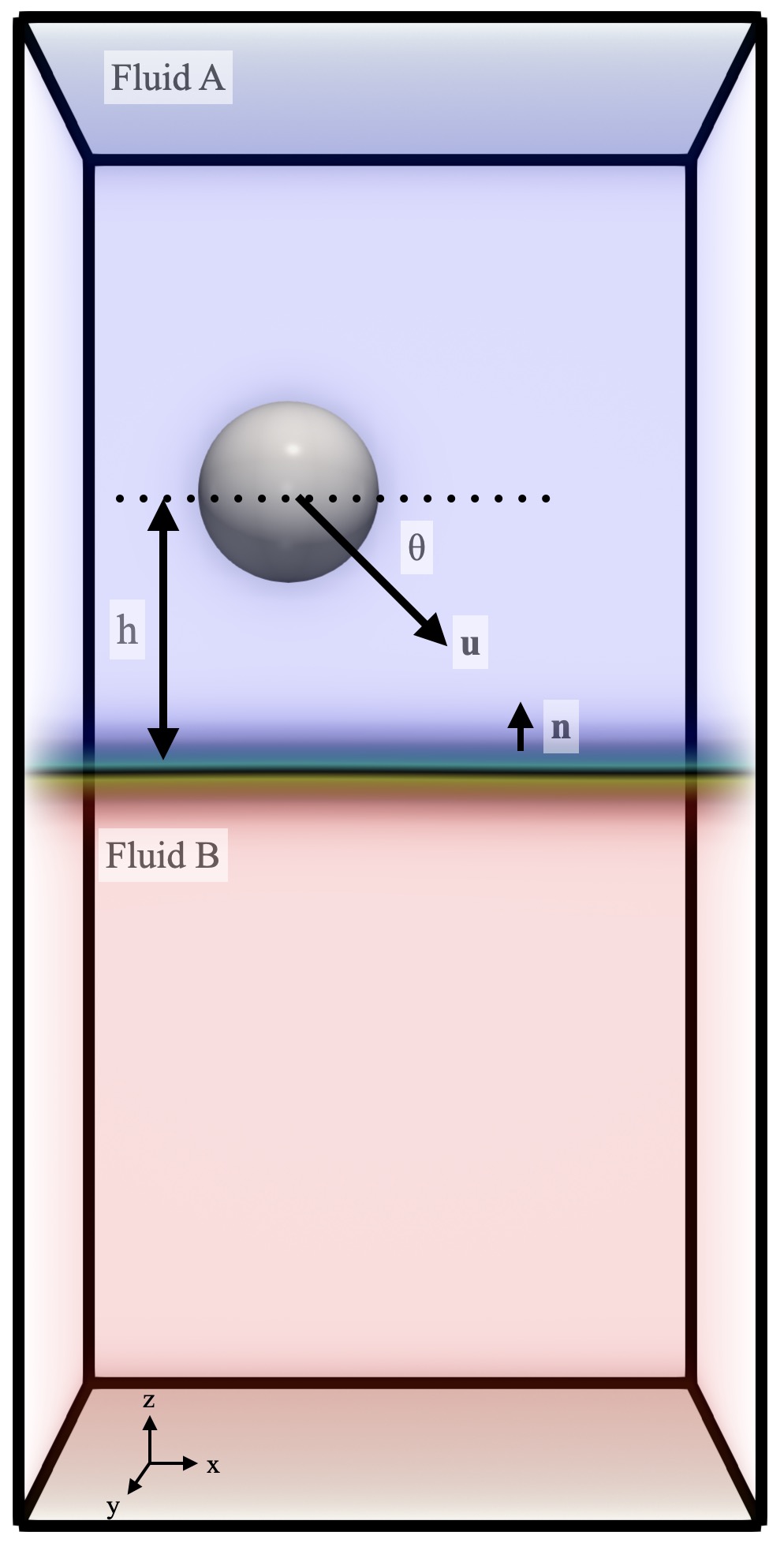}
\caption{\label{fig:schematic} Schematic of a single swimmer near an interface.}
\end{figure}

In this study, to investigate the dynamics of swimmers in inhomogeneous fluid systems, we focus on the dynamics of a single particle near a fluid--fluid interface. All simulations are conducted for an immiscible $A$/$B$ fluid system in a rectangular computational domain with dimensions of $32\Delta\times32\Delta\times64\Delta$, with $\Delta$ being the grid spacing and unit of length. Periodic boundary conditions are established in all directions.
Fluids $A$ and $B$ share all the same physical properties, such as density and viscosity, and are initially separated by phase in the $z$ direction (see Fig. \ref{fig:schematic}).

The radius of the squirmer is $a=4\Delta$.
In Eq.~\eqref{eq:freeenergy}, the parameter $\alpha$, which represents the coefficient of potential energy for the fluid A/B interface, is set to 1. The particle--fluid interface thickness $\xi_{p}$ and the fluid--fluid interface thickness $\xi_{f}$ are both set to $2\Delta$.
The parameter $B_1$ in Eq.~\eqref{eq:sq2} is set to 0.015, corresponding to a single-particle steady-state velocity of $U_0=2/3B_1=0.01$.
The mobility $\kappa$ (Eq.~\eqref{eq:cahn_hilliard}),
the shear viscosity $\eta$, and the mass densities $\rho=\rho_A=\rho_B=\rho_p$ are all set to $1$. Then, the particle Reynolds number is 
$Re=\rho{U_0}a/\eta=0.08$, 
the P\'eclet number is 
$Pe=U_0 a / \kappa=0.08$, and 
the Schmidt number is 
$Sc=Pe/Re=1$.

A schematic representation of our system is given in Fig.~\ref{fig:schematic}, which shows a single swimmer near a fluid--fluid interface. Thye deformable interfaces are initially planar and are located at $z=0$ and $z=L_z/2=32\Delta$. The distance between the center of mass of the swimmer and the nearest interface is $h$, with $h=16\Delta$ unless noted otherwise. The orientation of an interface is given by
its normal vector $\bm{n}$, and $\bm{u}$ denotes the swimmer's velocity. Since the initial velocity of the particle along the y-axis is set to 0, the swimmer will move only in the x--z plane. The angle $\theta=\arcsin{(\bm{n}\cdot{\bm{u}}/|\bm{u}|)}$ defines the orientation of the swimmer relative to the nearest interface.

\subsection{Motion near the interface}
To examine the motions of microswimmers near an interface, we conduct a series of simulations in which a swimmer approaches the nearest interface with different angles of approach $\theta_{in}{\in}({-\pi/2},\pi/2)$. The outgoing angle is denoted by $\theta_{out}$. The starting point of the swimmer is such that it will attain its steady-state velocity before it reaches the interface.

\begin{figure}[tbh]
    \includegraphics[width=0.95\linewidth]{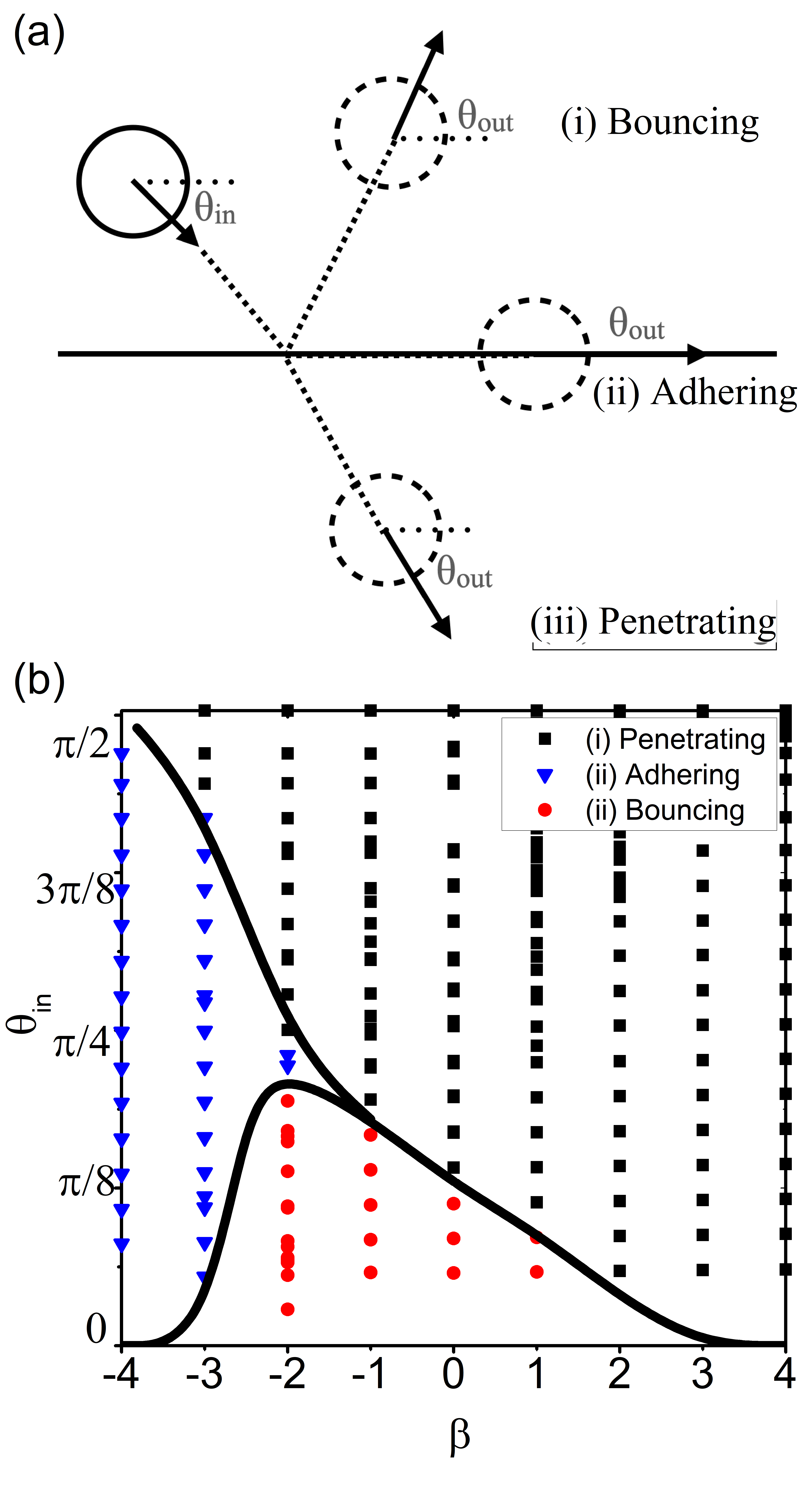}
\caption{\label{fig:motion_distri} (a) Sketchy of the three different swimming modes when a swimmer approaches an interface. (b) Diagram showing how these modes depend on the initial incidence angle $|\theta_{in}|$ and swimmer type $\beta$.}
\end{figure}

To understand the trajectories realised in our study consider a swimmer that starts off in host fluid A and approaches the interface with $\theta_{in}<0$. Three distinct ``collision'' modes are observed once the swimmer reaches the interface, namely, (i) ``bouncing'', (ii) ``adhering'', and (iii) ``penetrating'' motions, as illustrated in Fig.~\ref{fig:motion_distri}(a). In case (i), the swimmer bounces back into fluid $A$, avoiding fluid $B$, after performing a significant rotation within the interfacial domain and leaving the interface with $\theta_{out}>0$ (see movie S1 in the Supplemental Material). In case (ii), the swimmer becomes trapped at the interface, swimming in the x--y plane with $\theta_{out}=0$ (see movie S2 in the Supplemental Material). Finally, in case (iii), the swimmer passes through the interfacial barrier, swimming into fluid B with $\theta_{out}<0$ (see movie S3 in the Supplemental Material).

We conducted simulations with various initial angles $\theta_{in}$ and swimming parameters $\beta$ to construct a phase diagram for the three types of motions (i)--(iii), as shown in Fig.~\ref{fig:motion_distri}(b).
For weak swimmers, while the swimming strength and swimmer type play a role, the dominant factor determining the nature of the motion at the interface is the initial angle.
Generally, if $|\theta_{in}|$ is small, the swimmer will bounce back from the interface (i). If $|\theta_{in}|$ is large, the swimmer will swim across the interface (iii). For strong swimmers, strong pushers prefer to be absorbed at the interface with their swimming orientation aligned with the boundary (ii), while strong pullers are more inclined to cross the interface (iii).

\begin{figure*}
    \includegraphics[width=1\linewidth]{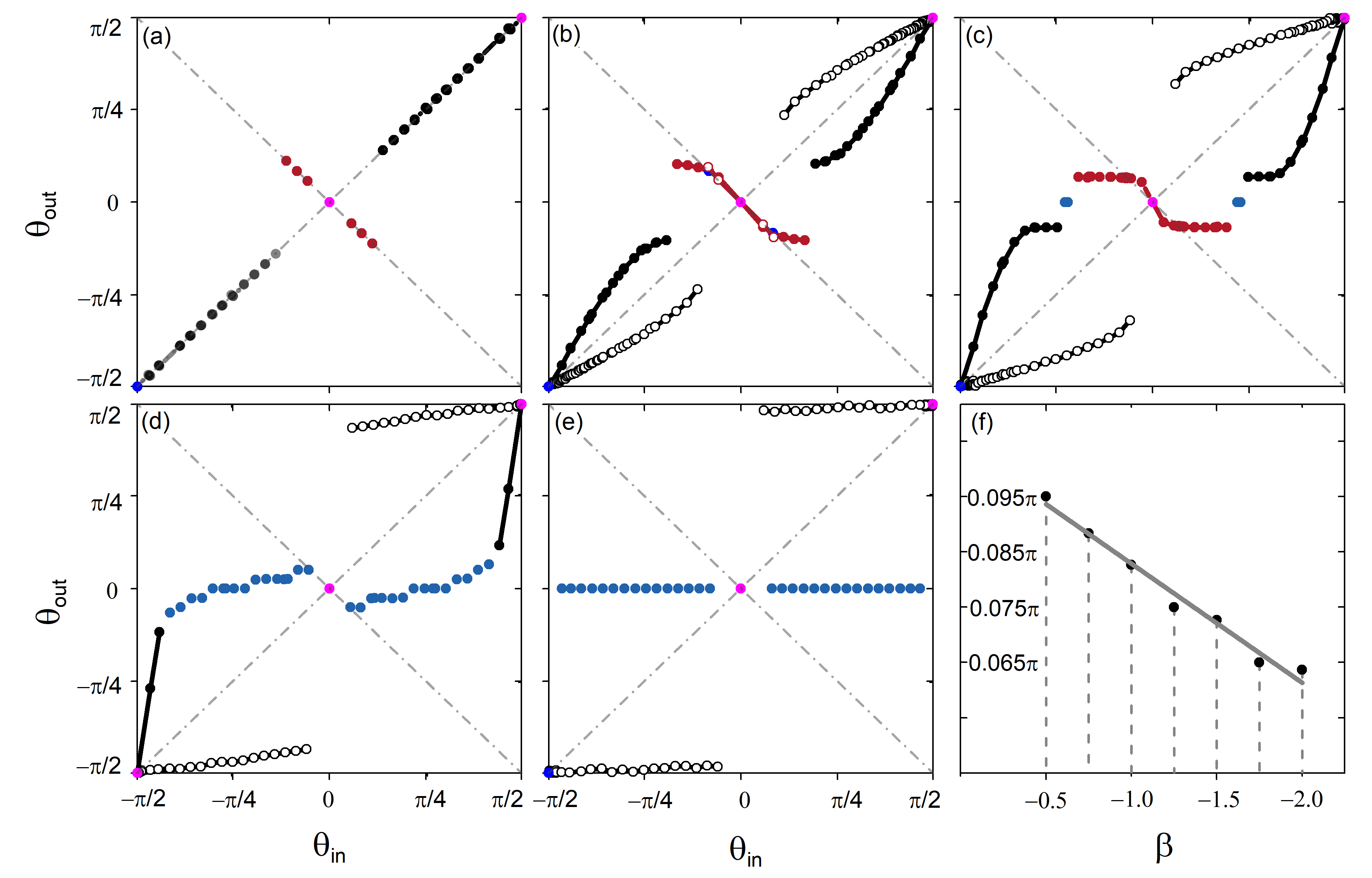}
\caption{\label{fig:change} Changes in the orientation angle $\theta$ for swimmers with various $\beta$ values: (a) $\beta=0$; (b) $\beta=\pm1$; (c) $\beta=\pm2$; (d) $\beta=\pm3$; (e) $\beta=\pm4$. Solid points represent pushers, while hollow points represent pullers. Black indicates crossing motion, and red indicates bouncing motion. Blue represents the special case in which the swimmer ultimately swims along the interface.
(f) Comparison of the final fixed angles $\theta^*$ for pushers with various $\beta$ values.}
\end{figure*}

The type and strength of the squirming behavior have a strong impact on the detailed dynamics of a swimmer at an interface, including the degree of rotation at the interface, i.e., the angle at which the swimmer leaves the interface.
To illustrate this effect, we conducted a large number of simulations with different initial angles for three types of swimmers ($\beta = -1, 0, 1$), as shown in Fig.~\ref{fig:change}(a)--(b).
For a neutral particle, the orientation angle $\theta$ shows no change after the swimmer leaves the interface, regardless of the swimming mode, as illustrated in Fig.~\ref{fig:change}(a).
However, pullers and pushers show a surprising symmetry in Fig.~\ref{fig:change}(b), especially for the penetrating mode.
Even if pushers and pullers start from the same initial angle $\theta_{in}$ and swim in the same mode (penetrating or bouncing), their orientation angles will change in different ways.
Taking the penetrating mode as an example, the angle $\theta_out$ between a pusher's orientation and the interface decreases after the swimmer leaves the interfacial domain, while a puller tends to swim perpendicular to the interface.

\begin{figure}[tbh]
    \includegraphics[width=1\linewidth]{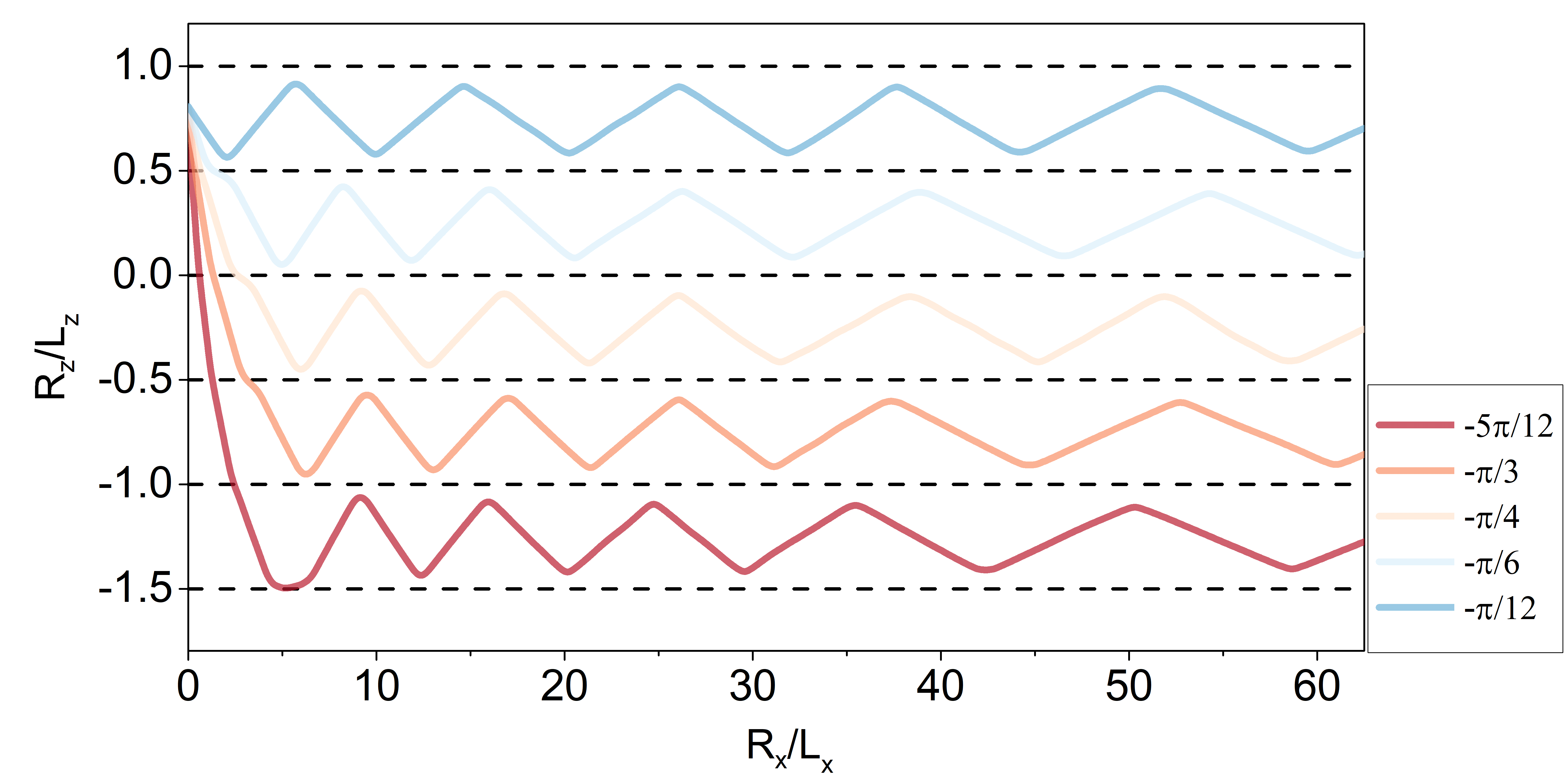}
\caption{\label{fig:trajectories}Swimmer ($\beta = -2$) trajectories showing repeated collisions with the interfaces (dashed lines) for various initial angles $\theta_{in}$ as shown in the key. The swimmer position is shown in units of the system height $L_z$ and width $L_x$ as it moves in $z$ (vertical) and $x$ (horizontal) respectively.}
\end{figure}

We also performed long-term simulations to study repeated collisions with interfaces. In the case of a pusher, a stable state of periodic back-and-forth motion between two interfaces is observed, as shown in Fig.~\ref{fig:trajectories}.
The steady-state motion is the same for all pushers, regardless of the initial angle.
For a puller,
the angle $\theta$ decreases during each pass, and finally, the swimmer reaches a steady state moving perpendicular to the interface.
Thus, we consider that the swimmer type has a strong effect on the dynamics near fluid--fluid interfaces.

\subsection{Swimmer types}

We first investigate the swimmer dynamics as a function of $\beta$.
As preliminary considerations, we assume that the following two boundary cases hold.
In the first case, when the initial angle is set to $0$, the swimmer will swim parallel to the interface due to the large separation from the interface.
In the second case, when the initial angle is set to $|\pi/2|$, the swimmer will swim perpendicular to the interface due to the symmetry of the system.
Therefore, by combining the above two boundary cases and simulation results, we can obtain a map $f$ relating the initial angle $\theta_{in}$ to the outgoing angle $\theta_{out}$ after the particle reaches the interface, as presented in Fig.~\ref{fig:change}.
The penetrating mode and the bouncing mode are represented in black and red, respectively.
For neutral particles, as shown in Fig.~\ref{fig:change}(a), the change in angle is insignificant.
Fig.~\ref{fig:change}(b)--(e) further show the differences between pushers and pullers with the same $|\beta|$ values ($1\leqslant|\beta|\leqslant4$).
Open symbols represent pullers, while filled symbols represent pushers.
From these graphs, we can observe that the maps for weak swimmers with opposite values of $\beta$ are nearly symmetric.
In general, swimmers with large initial angles, marked in black, swim across the interface.
On the other hand, swimmers with small initial angles, marked in red, bounce back from the interface.
However, the threshold angle that divides the bouncing and penetrating behaviors is different for different swimmers and depends on the $\beta$ value.
This is most easily seen from the penetrating trajectories of pullers and pushers, with the outgoing angle increasing for the former and decreasing for the latter. For pullers ($\beta>0$), these are the only two types of motion observed. For pushers ($\beta <0$), an additional ``adhering'' state is observed for $\beta \le -3$, marked in blue.

\begin{figure*}
    \includegraphics[width=1\linewidth]{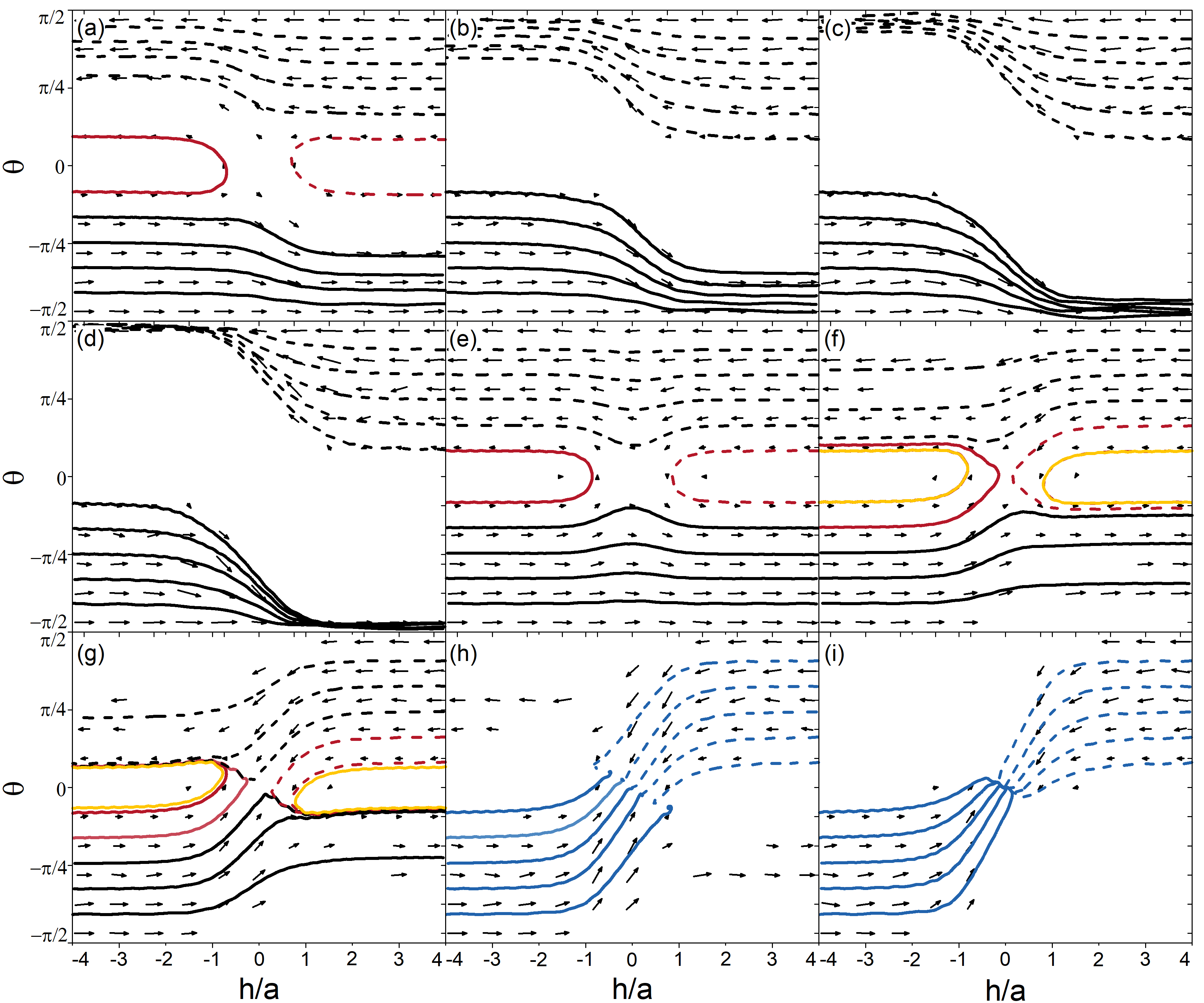}
\caption{\label{fig:SLVF} Vector field (arrows) showing the time evolution of the Orientation angle $\theta$ for swimmers with various $\beta$ values: (a)--(d) pullers with (a) $\beta=1$, (b) $\beta=2$, (c) $\beta=3$, (d) $\beta=4$; (e) neutral particle; (f)--(i) pushers with (f) $\beta=-1$, (g) $\beta=-2$, (h) $\beta=-3$, (i) $\beta=-4$. Black and red lines indicate crossing and bouncing-back motions, respectively, while yellow lines represent the steady states for the corresponding $\beta$ values and blue lines are used for swimmers that ultimately swim parallel to the interface. Solid and dashed flow lines indicate flows that are related by an inversion symmetry in $h$ and $\theta$.}
\end{figure*}

Fig.~\ref{fig:SLVF} shows how the orientation angle $\theta$ changes as a function of the distance from the nearest interface.
There is a clear asymmetry in the puller (a--d) and pusher (f--i) trajectories, which is not observed for neutral swimmers (e). Furthermore, for the case of pushers, the outgoing angle approaches a fixed value, with the swimmers reaching a steady state in which they bounce back periodically at this particular angle (marked in yellow in Fig.~\ref{fig:SLVF}). Once a pusher collides with an interface, it returns to the same position with the same orientation angle and then undergoes another collision. We will further discuss this steady state at the end of this section.

We note that the swimming strength also contributes to the hydrodynamic interactions near the interface. In particular, the change in orientation after crossing the boundary will be more pronounced for stronger swimmers. Thus, strong pullers will more quickly reach the stable state in which they swim perpendicular to the interface.
For pushers, for which the outgoing angle decreases, this can give rise to an adhering state. The corresponding trajectories are marked in blue in Fig.~\ref{fig:SLVF}(h) and (i).
In such a case, the pusher can move along the interface, with half of its body in fluid A and the other half in fluid B.
This motion is reminiscent of the equatorial anchoring of Janus particles at an oil--water interface \cite{park2011janus}.
However, the former is due to the symmetry of the fluid system about the interface, while the latter is due to the symmetrical structure of the amphiphilic particles.
Additionally, according to Fig.~\ref{fig:motion_distri}(b), the range of initial angles that can lead to this adhering state increases as the pusher becomes stronger.

\begin{figure}[tbh]
    \includegraphics[width=1\linewidth]{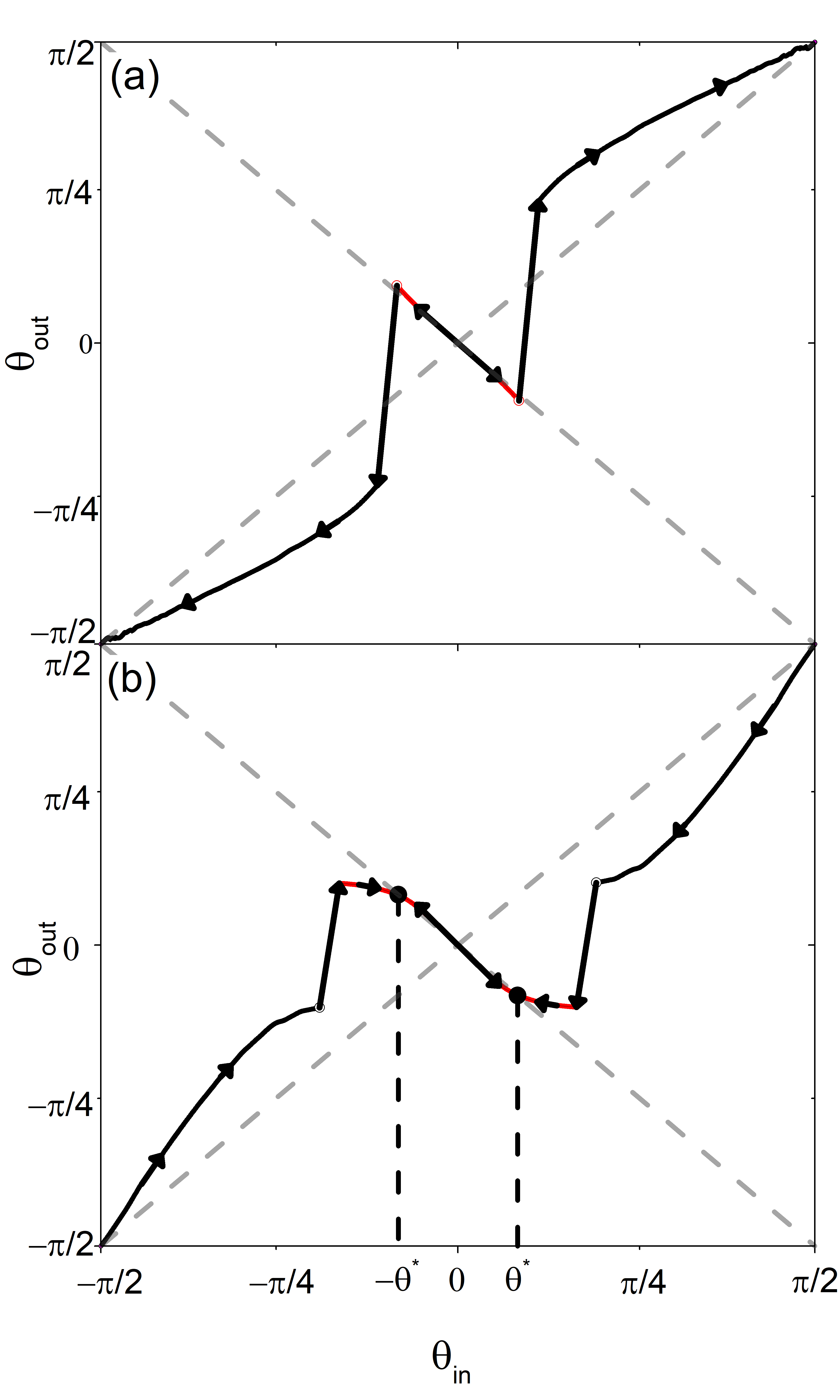}
\caption{\label{fig:finalstate} Trends in the orientation angle $\theta$ for (a) pullers ($\beta=1$) and (b) pushers ($\beta=-1$).}
\end{figure}
Due to the 
symmetry considered in this work, i.e. alternating fluid layers with identical properties for the two fluids, the swimmer trajectories show convergence after several interfacial interactions.
According to Fig.~\ref{fig:finalstate}(a), the terminal angle of a weak puller will eventually converge to either $\pi/2$ or $-\pi/2$, regardless of the initial angle.
That is, after it has repeated the process of approaching an interface several times, a puller will eventually swim perpendicular to the interface, as shown by the dashed-line trajectory in Fig.~\ref{fig:final} (see movie S4 in the Supplemental Material).

\begin{figure}[tbh]
    \includegraphics[width=1\linewidth]{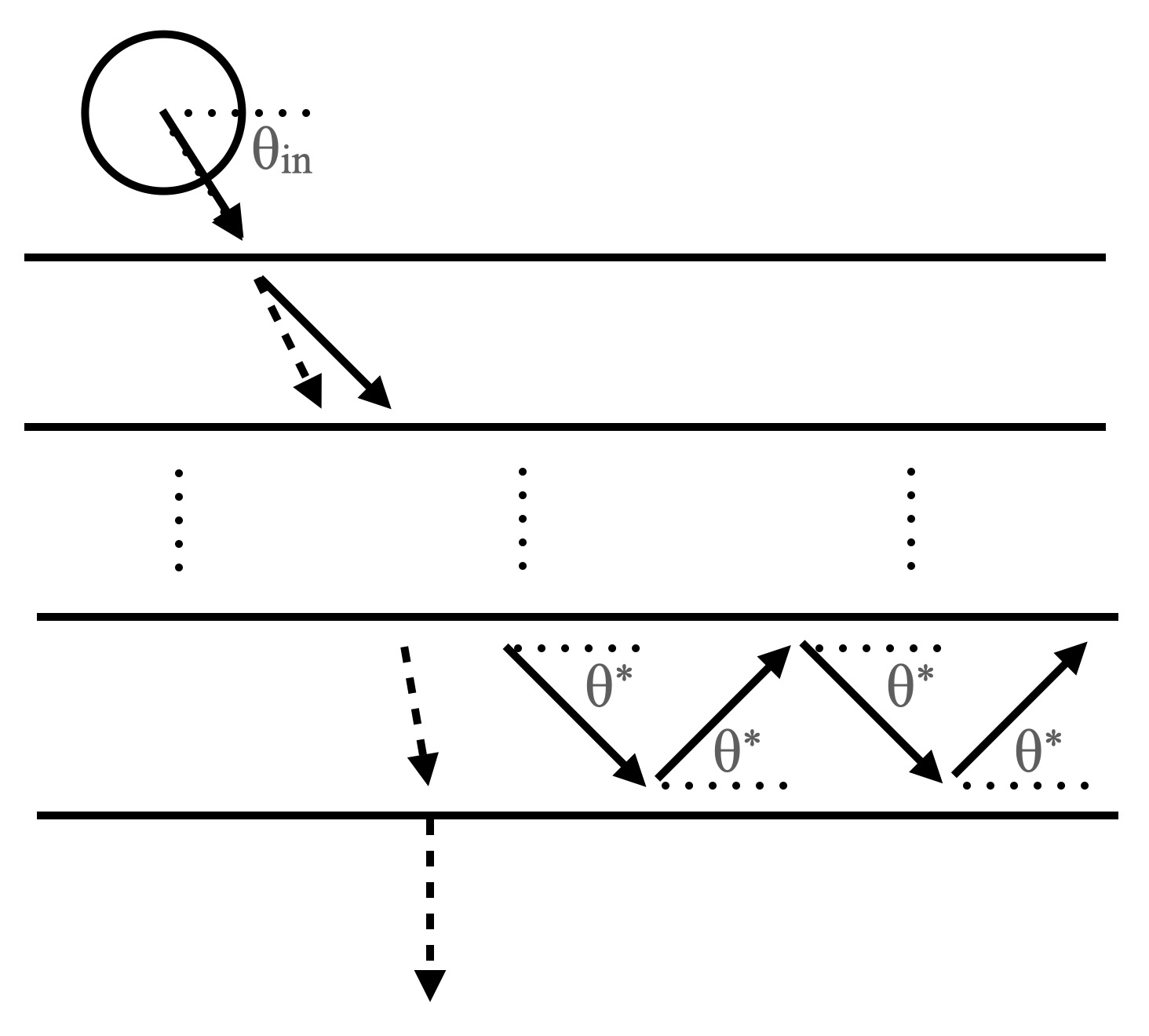}
\caption{\label{fig:final} Two types of stable states for swimmers. Trajectory represented by dashed lines: A puller eventually swims perpendicular to the interface. Trajectory represented by solid lines: A pusher eventually bounces back and forth between two interfaces while remaining in one fluid domain. }
\end{figure}

In addition, for initial angles other than the boundary cases of $\theta=\pm \pi/2$, the terminal angle for a pusher will eventually converge (after repeated interfacial collisions) to an intersection point $\theta^*$
that is located in the bouncing motion regime, as shown in Fig.~\ref{fig:finalstate}(b).
That is, pushers will always stabilize to a state in which they bounce back and forth at a fixed angle $\theta^*$, as shown by the solid-line trajectory in Fig.~\ref{fig:final} (see movie S5 in the Supplemental Material).
This fixed angle $\theta^*$ depends on the value of $\beta$, as illustrated in Fig.~\ref{fig:change}(f). For sufficiently strong pushers, swimming along the interface is also a possible steady state, marked in blue in Fig.~\ref{fig:change}.

\section{discussion}
\begin{figure}[tbh]
    \includegraphics[width=1\linewidth]{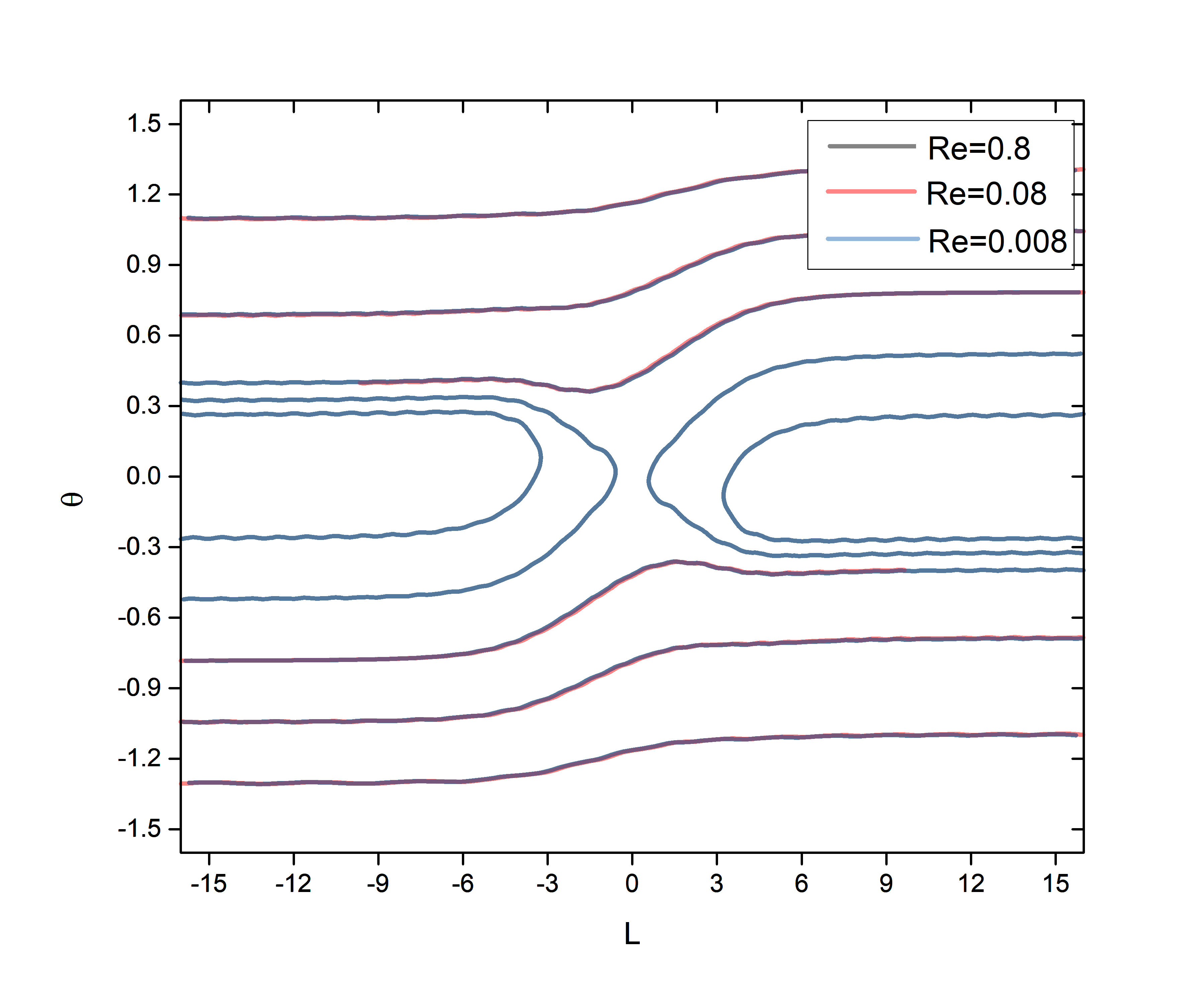}
\caption{\label{fig:Re}Variation in the orientation angle $\theta$ with (a) a fixed Reynolds number (Re), (b) a fixed Schmidt number (Sc), and (c) a fixed P\'eclet number (Pe). }
\end{figure}

All the simulations discussed above were conducted for a fixed Reynolds number (relative swimmer speed normalized with momentum transport rate), a fixed P\'eclet number (relative swimmer speed normalized with $\psi$ transport rate), and a fixed Schmidt number (relative momentum transport rate normalized with $\psi$ transport rate), which were set to $Re=0.08$, $Pe=0.08$, and $Sc=1$, respectively, meaning that inertial effects are expected to be negligible.
To examine the contribution of inertial effects to the swimmer dynamics, we also conducted some additional simulations for different values of $Re$ and $Pe$.
Fig.~\ref{fig:Re} shows the variation in the orientation angle with the distance of the swimmer from the interface.
The parameters used are the same as those in Fig.~\ref{fig:SLVF}(f), except for the values of $U_0$.
We compare three cases of pushers with $Re=Pe=0.008$, $0.08$, and $0.8$ in Fig.~\ref{fig:Re}, where it is seen that the three trajectories perfectly agree with each other.
This result indicates that the inertial effects are negligible in our present simulations.
Although the effect of $Sc$ is not considered in the present study, it is also likely to contribute to the swimmer's dynamics as it approaches the interface. The exact mechanisms for this will require further investigation. 

\color{black}
To the best of our knowledge, previous numerical studies of swimmer dynamics at interfaces \cite{ishimoto2013squirmer,pimponi2016hydrodynamics} have
usually considered only far-field hydrodynamics or non-penetrable surfaces. 
However, in this paper, we focus on a physical model in which the boundary is a soft, deformable and penetrable interface rather than simply being defined as impassable. Thus, our model also accounts for the interaction of swimmers with such a penetrable fluid--fluid interface. As a result, novel dynamics can be predicted and analysed, such as the penetrating mode.

The mode in which the swimmer adheres at the interface has previously been reported by Deng et al \cite{deng2020motile} who observed that \textit{Pseudomonas aeruginosa} adsorbed onto an oil--water interface and swam in one of four characteristic motility modes which they term visitor, diffusive, pirouette, or curly.

The adhering state was previously studied by means of a general multipole-expansion-based singularity model for swimming microorganisms \cite{desai2020biofilms}.
Both pushers and pullers were predicted to accumulate at an oil--water interface, giving rise to large density inhomogeneities in many-particle systems. The collective dynamics of microswimmers strongly affects their motion \cite{elgeti2015physics,gompper2016microswimmers}.
They can exhibit highly organized movements with remarkable large-scale patterns, such as networks, complex vortices, or swarms. In the present work we analyse only a single swimmer. This might help to explain why we predict instead that only strong pushers can be trapped by an interface. 
Li and Ardekani's work \cite{li2014hydrodynamic} is probably the closest in methodology to  work, although they studied the motion of microswimmers near a solid wall. They found that a swimmer that was initially oriented toward the wall can escape (bounce back) if the strength of its squirming is sufficiently weak.
However, they also reported another swimming mode, in which very strong swimmers ($|\beta|>7$) were observed to repeatedly bounce at the wall, which we do not observe in our simulations of a soft interface, although a harder interface would be accessible within our methodology.

\section{Conclusions}

In this paper we analyse the dynamics of microswimmers in a binary fluid system.
Our simulations are based on the smoothed profile method and the squirmer model. This allows accurate and efficient analysis of the dynamics near deformable fluid--fluid interfaces.
Three qualitatively distinct dynamical modes emerge for swimmers approaching an interface, (i) crossing, (ii) adhering, and (iii) bouncing. The dynamical properties depend on the swimmer type, the swimming strength, and the initial angle of approach.
For a puller, the orientation angle is predicted to increase after the swimmer interacts with the interface. This will eventually reach $\pm \pi/2$ after repeated interfacial collisions after which the puller will swim perpendicular to the interface.
For a pusher the orientation angle instead approaches a fixed oblique angle $\theta^*$ by increasing or decreasing, depending on whether the initial orientation was smaller or greater than this angle respectively.
As a consequence of this, we observe that most pushers will eventually exhibit a steady-state mode in which they bounce between two interfaces along trajectories inclined at angle $\theta^*$. This steady-state angle $\theta^*$ is related to the swimmer type. The other possible dynamical mode arises for the case of a strong pusher, for which swimming parallel to the interface emerges as another possible steady state.

Our results provide a detailed analysis of the hydrodynamic interactions of microswimmers with a deformable fluid--fluid interface. This improves our understanding of microswimmer motion in environments involving soft interfaces, having some similarity with those found in Biology. Our study may also have some relevance in the context of bioengineering applications. For example, we could also incorporate additional features into our model, such as the nutrient chemotaxis.
\color{black}

\begin{acknowledgments}
The authors express their gratitude to Dr. Hiroto Ozaki and Dr. Takeshi Aoyagi for their collaboration and outstanding contributions to the simulation software development.
R.Y. acknowledges helpful discussions with Profs. Hajime Tanaka and Akira Furukawa.
This work was supported by the Grants-in-Aid for Scientific Research (JSPS KAKENHI) under grant nos. JP 20H00129, 20H05619, and 20K03786 and by the NEDO Project (JPNP16010).
\end{acknowledgments}

\bibliography{ref.bib}

\clearpage
\begin{widetext}
\setcounter{page}{1}

\begin{center}
Supplemental Information for the Manuscript \\
``Dynamics of microswimmers near a soft penetrable interface''\\
\ \\
Chao Feng, John J. Molina, Matthew S. Turner, and Ryoichi Yamamoto\\
Department of Chemical Engineering, Kyoto University, Kyoto 615-8510, Japan
\end{center}

\ \\
\ \\
\ \\

\noindent {\bf Supplementary Movie 1}\\
\\
{\bf Description:} Movie of a puller-type swimmer with $\beta=3$ at an orientation angle of $\theta= 5\pi/6$, where the swimmer penetrates through the fluid--fluid interface.
\\

\noindent {\bf Supplementary Movie 2}\\
\\
{\bf Description:} Movie of a pusher-type swimmer with $\beta=-4$ at an orientation angle of $\theta= 5\pi/6$, where the swimmer becomes adhered to the fluid--fluid interface.
\\

\noindent {\bf Supplementary Movie 3}\\
\\
{\bf Description:} Movie of a weak puller-type swimmer with $\beta=-1$ at an orientation angle of $\theta= 5\pi/6$, where the swimmer bounces back from the fluid--fluid interface.
\\

\noindent {\bf Supplementary Movie 4}\\
\\
{\bf Description:} Identical to Supplementary Movie 1 but from a longer-term simulation. The swimmer reaches a steady state in which it repeatedly penetrates through the fluid--fluid interface (see Fig.~7).
\\

\noindent {\bf Supplementary Movie 5}\\
\\
{\bf Description:} Identical to Supplementary Movie 3 but from a longer-term simulation. The swimmer reaches a steady state in which it repeatedly bounces back from two adjacent fluid--fluid interfaces (see Fig.~7).


\end{widetext}

\end{document}